# NUMERICAL SIMULATION AND ANALYSIS OF THERMALLY EXCITED WAVES IN PLASMA CRYSTALS

K. Qiao and T. W. Hyde

*CASPER, (Center for Astrophysics, Space Physics and Engineering Research)*
*Baylor University, Waco, TX. 76798-7310, USA*

**ABSTRACT**

A numerical model for a 2D-monolayer plasma crystal was established using the Box_tree code. Box_tree is a Barnes_Hut tree code which has proven effective in modeling systems composed of large numbers of particles. Thermally excited waves in this plasma crystal were numerically simulated and dispersion relations for both the longitudinal and transverse wave modes were found. These were compared with the dispersion relations extrapolated from experiment as well as a theory based on harmonic approximation. The results were found to agree with theoretical dispersion relations under different wave propagation directions with different particle charges and over a range of $0.9 < \kappa < 5$.

**INTRODUCTION**

Dusty plasma systems play an important role both in astrophysical environments (for example, proto-stellar clouds and ring systems) and laboratory situations (plasma processing and nanofabrication). In general, there is still little or no evidence of strongly coupled dusty plasmas in nature. However, over the past two decades it has been shown that dust particles immersed in a plasma or radiation environment collect charges with their subsequent dynamics becoming influenced - or in some cases dominated - by the local electric and magnetic fields (Matthews and Hyde, 1998). (This is in addition to the usual mechanisms which modify grain dynamics such as the gravitational, radiation pressure or gas drag effects.) There are a number of unusual observations – such as the spokes at Saturn or the dust streams escaping Jupiter - that can still only be understood through recognizing dusty plasma effects.

In addition, the formation and stability mechanisms for ordered colloidal (Coulomb) crystals within a tenuous dusty plasma is of great interest in protoplanetary, protostellar and accretion disk formation as well as spiral galaxies and dark matter research. Ever since crystal formation within dusty plasmas was discovered experimentally at the Max Plank Institute by Thomas, et al. (1994), interest in plasma crystals has increased dramatically.

Recent research has focussed (both theoretically and experimentally) on the different types of wave mode propagation which are possible within plasma crystals. This is an important topic since several of the fundamental quantities for characterizing such crystals can be obtained from an analysis of the wave propagation/dispersion.

Waves in 2D plasma crystals for both longitudinal and transverse modes have been observed in several recent experiments (Homann et al., 1997 and 1998, Nunomura et al., 2000 and 2002a). In all of these, the 2D plasma crystals are initially formed by levitating polymer microspheres in a rf discharge plasma. The particles are negatively charged by the plasma and interact with each other through a screened Coulomb repulsion

$$U(r) = Q(4\pi\varepsilon_0 r)^{-1} \exp(-r/\lambda_D),$$

where $\lambda_D$ is the Debye length. Particles are levitated in the vertical direction by the DC self-bias and trapped in the horizontal direction by an electric field in the sheath region above the powered electrode with the rarefied neutral gas in the plasma damping any collective particle motion. Due to the screened Coulomb interaction between the particles and the neutral gas drag which limits the particle's kinetic energy, the particles form an ordered lattice in the horizontal direction. Generally, the lattices formed are determined to be a monolayer hexagonal lattice characterized by a screening parameter $\kappa = a/\lambda_D$, where a is the interparticle spacing.

In some experiments, the observed waves were produced by exciting the lattice intentionally. Modulated laser beams produced a sinusoidal waveform which was then used to perturb particles in a specific region of the lattice. The subsequent waves produced propagated away from the region and were analyzed, yielding dispersion relations. Two wave modes, longitudinal (Homann et al., 1997 and 1998, Nunomura et al., 2002a) and transverse (Nunomura et al., 2000), were found having different dispersion relations. The dispersion properties of the waves, especially the transverse mode, were found to be dependent on the direction of propagation relative to the lattice. The dispersion relations evaluated in this manner were shown to agree with the theoretical model of Wang et al. (2001).

**SIMULATION MODELS AND METHODS**

In the past, dispersion relations determined experimentally using laser excitation of waves such as those described above, have been unable to show the negative group speed predicted by theory and the wave experiences heavy damping at high frequencies. However, recently, an experiment verified the theoretical dispersion relations to the first Brillouin zone by detecting the random particle motion within the lattice (Nunomura et al. 2002). In this work, the Box_tree code was used to simulate such a thermally excited (spontaneous) wave within a plasma crystal. Dispersion relations were determined using the data and then analyzed for different parameters, including the particle charge and Debye length.

The Box_Tree code is a Barnes-Hut tree code first written by Derek Richardson (1994) for planetary ring and disk studies. It was later modified by Lorin Matthews (1998) to include electrostatic interactions and then by John Vasut (2001) to simulate the formation of plasma crystals. It has proven to be an effective tool for modeling real time systems composed of large numbers of particles with specific interparticle interactions. The Box_Tree code models a dusty plasma system by first dividing it into self-similar patches, where the box size is much greater than the radial mean excursions of the constituent dust grains and the boundary conditions are met using twenty-six ghost boxes. A tree code is incorporated into the Box_Tree routine to allow it to deal with interparticle interactions. Since most such interactions can be determined by examining the multipole expansion of collections of particles, the code scales as N•logN instead of $N^2$, resulting in greater CPU time efficiency. Data files showing each particle's position and velocity are output for analysis after a user defined time interval. Particle charge, mass, density, Debye length and output data intervals are all adjustable parameters in the code.
Fig. 1. Dispersion relation from the simulation of (a) longitudinal and (b) transverse waves in a coulomb crystal. ($\kappa$ = 1.4 ($\lambda_D$ = 570 μm), q =1.92*$10^{-15}$ C.)

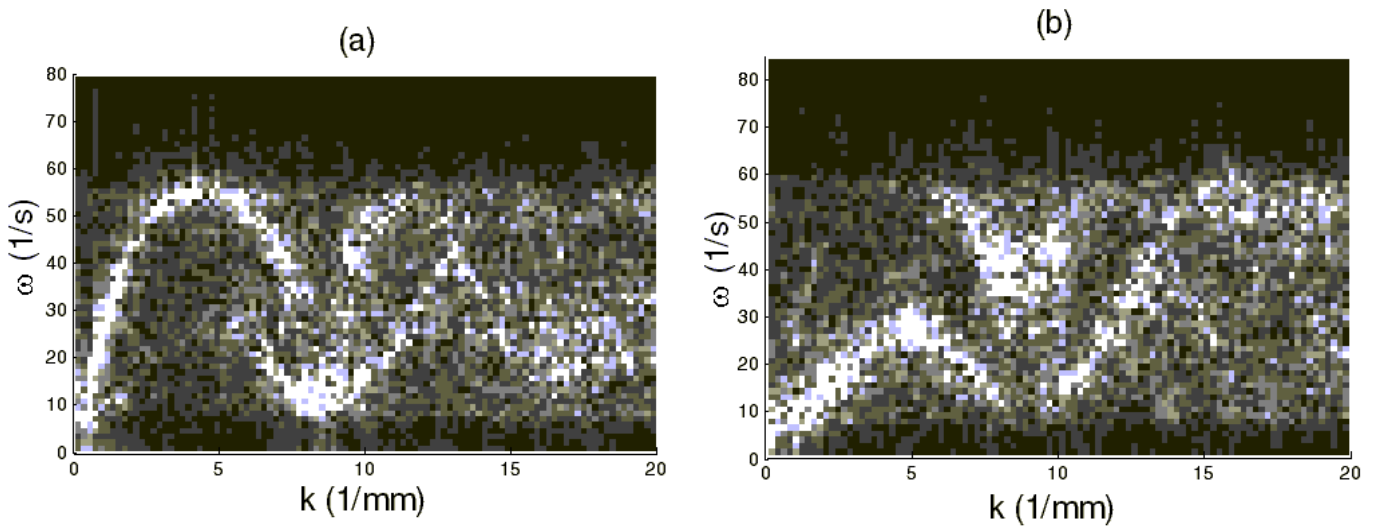

The interparticle interactions include gravitational and electrostatic forces. The electrostatic force is assumed to be a 3D screened Coulomb repulsion or Yukawa potential. The simulation box is a 3D box and all calculations are handled in a 3D manner. 3D periodic boundary conditions are met using twenty-six ghost boxes as mentioned above. However, the initial positions of the particles are constrained to lie in a horizontal plane (as is often the case in an experimental setting) and their initial velocities are specified to contain only XY components. The gravitational force from the earth and the electrostatic force produced by the (lower) powered electrode are not assumed to balance one another and are neglected. Therefore, the particles are restricted to a 2D plane (within the 3D box) throughout the simulation since they are not acted upon by forces (or velocities) in the Z direction.

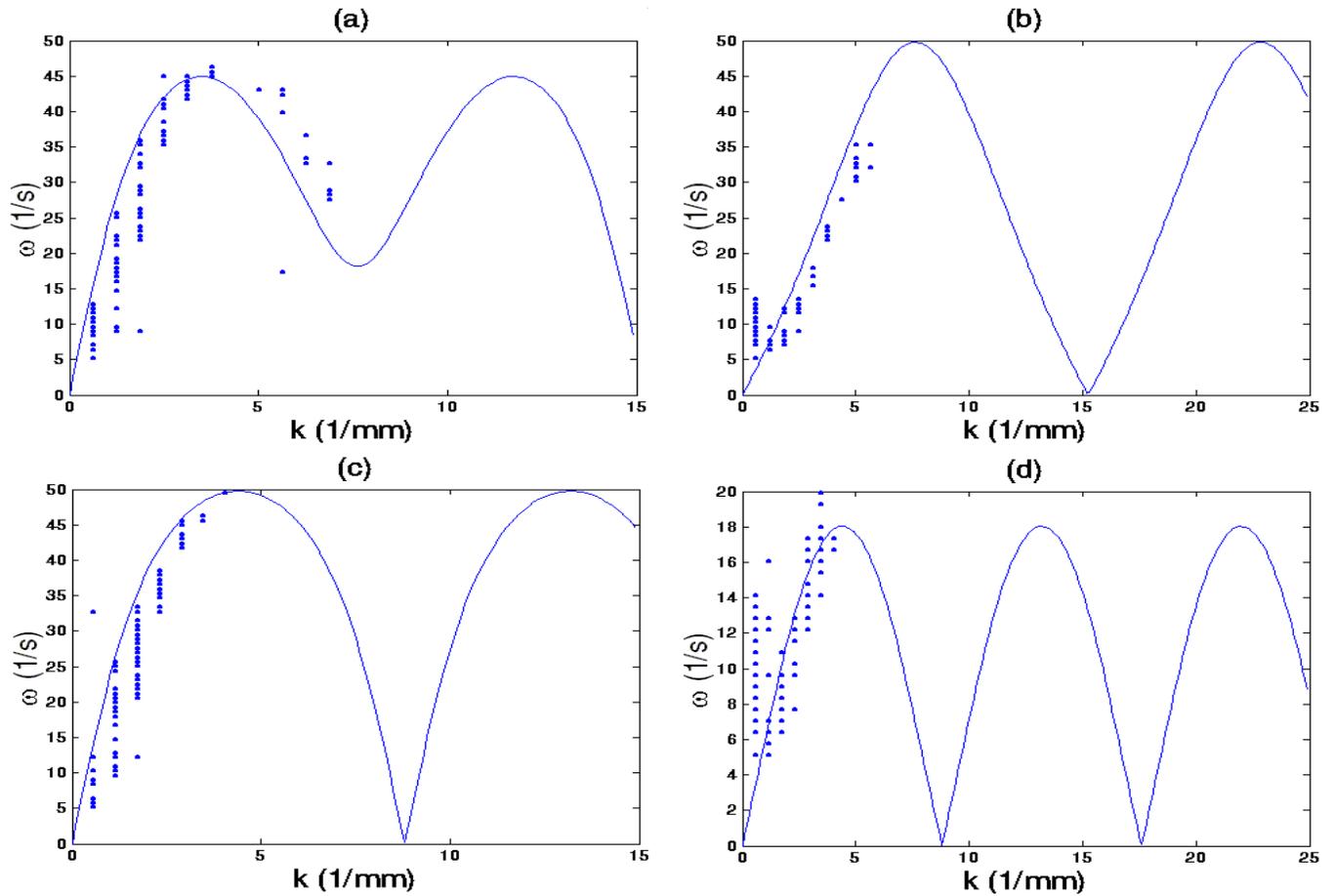

Fig. 2. Dispersion relations of (a) longitudinal and (b) transverse waves propagating parallel to the primitive translation vectors, and of (c) longitudinal and (d) transverse wave propagating perpendicular to the primitive translation vectors. $\kappa = 1.4$ ($\lambda_D = 570$ μm), $q = 1.92 \times 10^{-15}$ C.

Initial conditions include a random distribution of 679 particles in a $20 \times 20 \times 20$ mm$^3$ box. The neutral gas drag is included with a specified Epstein drag coefficient of $\gamma = 1.22$. After about 20 seconds, the state of the system was determined to be in a crystalline form (solid) using the pair correlation function. The lattice was determined to be triangular (hexagonal) using the Voronoi diagram (Vasut and Hyde, 2001). In all of the simulations, the diameter of the embedded particles is 6.5 μm, the density is 1.51 g/cm$^3$, the charge is 14,500 e and the interparticle distance as determined from the pair correlation function is around 825 μm. The initial position of the particles are generated by Box_tree in a random manner subject to the condition that the system's center of mass should be located at the center of the box. The initial velocities of the particles are generated using a Gaussian distribution with a mean of zero in both the x and y directions. The size of the standard deviation as measured from zero is supplied as a run-time parameter. As time goes on, particle velocities are simply calculated from former particle positions, velocities and interparticle forces. System temperatures can be determined from the average kinetic energies (velocities).

Once the system is shown to be in the form of a solid crystal, subsequent particle data, including position and velocity, are output using a time interval of 0.01 seconds and collecting 1000 data files for each simulation. Depending on the equilibrium particle position, the particles are divided into bins with the particle velocity averaged over each bin, yielding velocity data which is dependent upon position. Combining data files, a velocity matrix depending on time and position is obtained with a double Fourier transformation of this matrix yielding particle velocities in ω-k space. Figure 1 shows this matrix with pixel brightness corresponding to velocity value. The bright areas show the dispersion relation of the spontaneous wave for $\kappa = 1.4$ ($\lambda_D = 570$ μm) and particle charge, q =1.92*10$^{-15}$ C. To obtain the dispersion relation for a particular wave mode (longitudinal or transverse) with a specific propagation direction, the bins have to be chosen perpendicular to the propagation direction and velocity components perpendicular to the bins for longitudinal waves and parallel to the bins for transverse waves must be used, i.e., the velocities sampled are thus parallel to the propagation direction for the longitudinal mode and perpendicular to the propagation direction for the transverse mode.
Fig. 3.  Same as Figure 2 but q =2.72*10$^{-15}$ C.

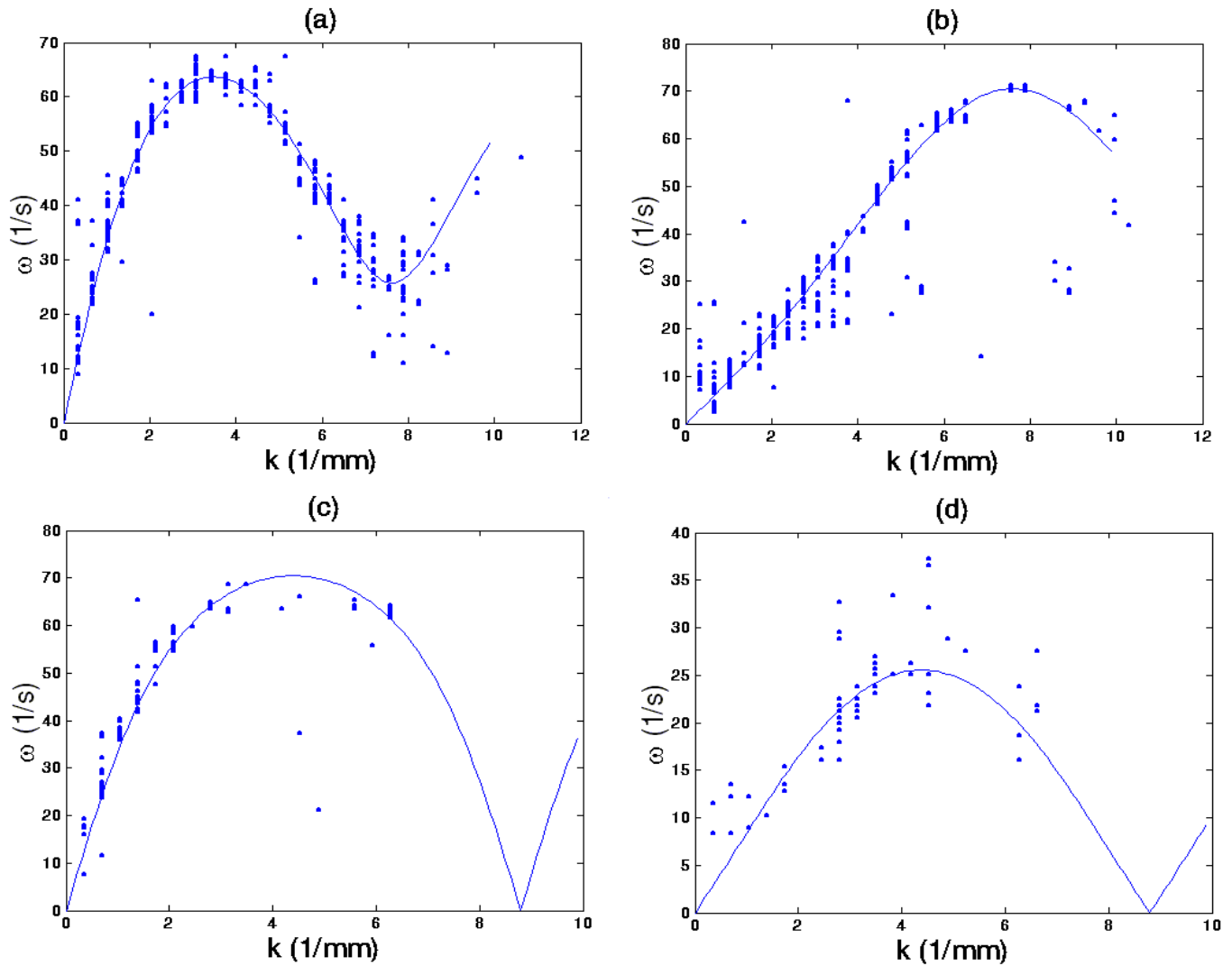

The possibility of temperature fluctuations of the system (temperature control / system thermostat) is carefully monitored and has been shown to be negligible for all conditions examined by this simulation (Vasut and Hyde, 2001). The dispersion relations were also examined for various temperatures to ensure that they did not depend on the overall system temperature.

**RESULTS**

X. Wang et al. (2001) analytically derived the dispersion relations for both longitudinal and transverse waves in a 2D perfectly triangular lattice with the primitive lattice vectors all in the same direction. He also assumes a Yukawa potential between the particles and a harmonic approximation of the motion of the particles about their equilibrium positions. The resulting dispersion relations were shown to depend upon the direction of propagation. For both modes, the dispersion relations for waves propagating parallel to the primitive translation vectors are different from those perpendicular to the primitive translation vectors (Figure 2) as has been shown by experiment (Nunomura, 2002b). In this case, the dispersion relations for both wave modes and propagation directions were obtained by simulation and analysis, as shown in Figure 2. Superimposing the dispersion relation obtained from Wang's theory as a solid line over the data allows a fit between the two to be determined. As can be seen in Figures 2 (b) and (d), the dispersion relation is very different for the transverse mode for the two propagation directions, while the longitudinal mode has nearly the same dispersion relation for the two directions, as can be seen in Figures 2 (a) and (c).

The simulation was run for varying κ values over the regime 0.9<κ<5. Dispersion relations for all values within this range were found to agree with theoretical curves (Qiao and Hyde, 2003). For κ > 5, the dispersion curve is too low (and the shape is no longer clear enough) to obtain distinguishable data. On the other hand, when κ is too small, the system will not form a crystal lattice and instead will remain in a liquid form.
Fig. 4. Same as Figure 2 but q =1.52*10$^{-15}$ C.

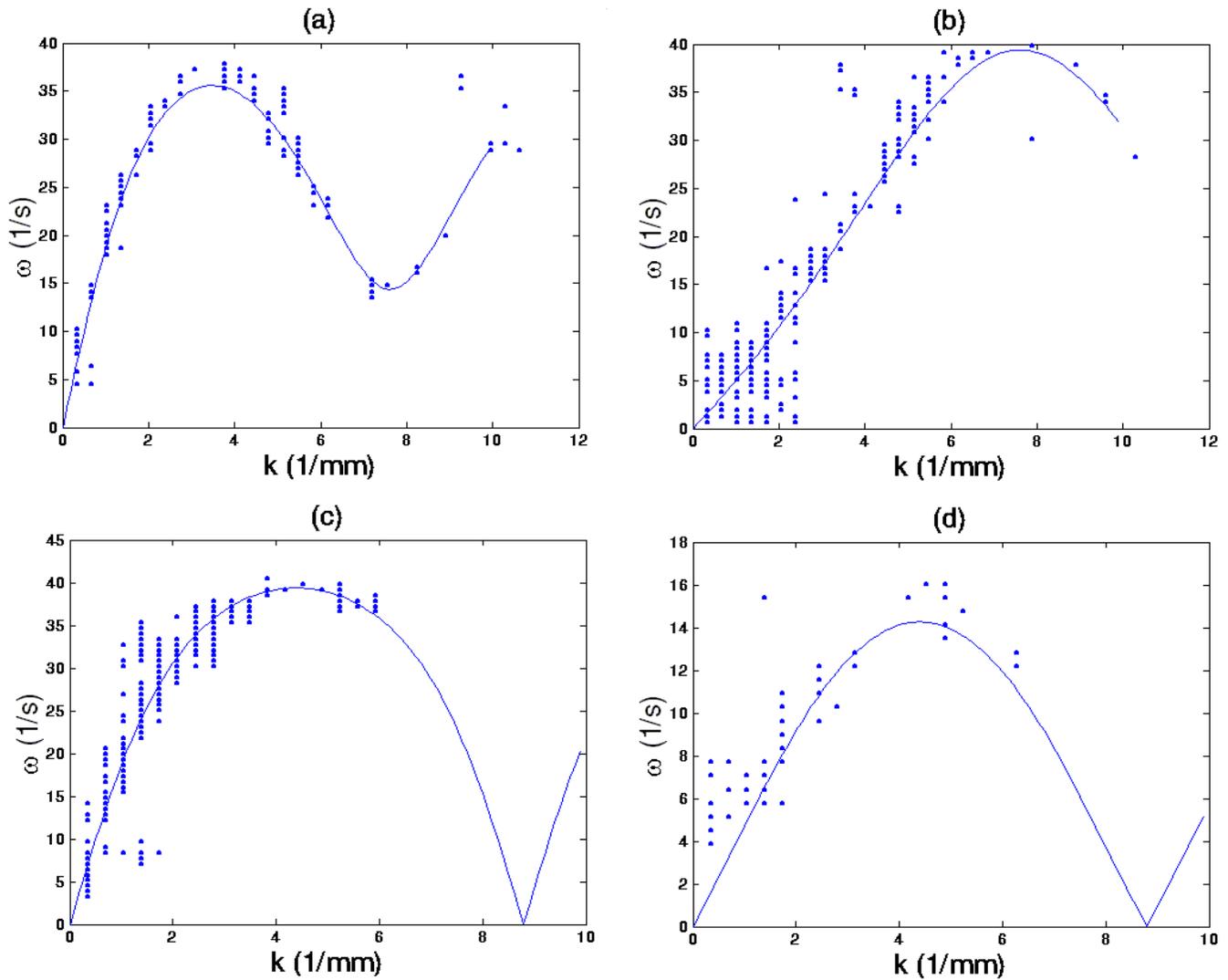

Additionally, the simulation was run for two other different particle charges, q = 1.52*10$^{-15}$C and q = 2.72*10$^{15}$C. Dispersion relations for longitudinal and transverse wave modes under different wave propagation directions for both charge values are again found in agreement with the theoretical curves, as shown in Figure 3 and Figure 4.

# CONCLUSIONS

As can be seen, Box_tree produces dispersion relations shown to agree with theory over a larger range of κ than experimentally possible. Additionally, the code allows adjustment of the fundamental parameters in a much easier manner than does experiment and has no limit on what parameters can be examined. (For example, the code can easily cover high frequencies when needed.) Finally, the only assumptions made in the program are a Yukawa potential and the assignment of various constant parameters (such as particle charge, mass or Debye length). This will allow it to be effective in simulating nonlinear effects and dusty liquids.

# REFERENCES

Homann, A., A. Melzer, S. Peters et al., Determination of the dust screening length by laser-excited lattice waves, *Phys. Rev. E,* **56**, 7138-7141, 1997.
Homann, A., A. Melzer, S. Peters et al., Laser-excited dust lattice waves in plasma crystals, *Physics Letters A,* **242**, 173-180, 1998.
Thomas, H., G.E. Morfill, V. Demmel et al., Plasma crystal: coulomb crystallization in a dusty plasma, *Phys. Rev. Letters*, **73**, 652-655, 1994.
Richardson, D. C., A new tree code method for simulation of planetesimal dynamics, *Mon. Not. R. Astron. Soc*, **261**, 396-414, 1993.
Vasut, J. and T. Hyde, Computer simulations of coulomb crystallization in a dusty plasma, *IEEE transactions on plasma science* **29**, 231-237, 2001.
Matthews, L. and T.W. Hyde, Numerical simulation of gravitoelectrodynamics in dusty plasmas, in *Strongly Coupled Coulomb Systems,* edited by Kalman et al., pp.199-202, Plenum Press, New York, 1998.
Nunomura, S., D. Samsonov, J. Goree, et al., Transverse waves in a two-dimensional screened-Coulomb crystal (dusty plasma), *Phys. Rev. Letters* **84**, 5141, 2000.
Nunomura, S., J. Goree, S. Hu, et al., Dispersion relations of longitudinal and transverse waves in two-dimensional screened Coulomb crystals, *Phys. Rev. E* **65**, 66402, 2002a.
Nunomura, S., J. Goree, S. Hu, et al., Phonon spectrum in a plasma crystal, *Phys. Rev. Letters* **89**, 35001, 2002b.
Wang, X., A. Bhattacharjee, S. Hu et al., Longitudinal and transverse waves in Yukawa crystals, *Phys. Rev. Letters* **86**, 2569, 2001.
Qiao, K. and T. Hyde, Dispersion relations for thermally excited waves in plasma crystals, in press, *J. Phys. A*, 2003.

Ke_Qiao@Baylor.edu, Truell_Hyde@Baylor.edu